\documentclass[11pt,onecolumn,amssymb,nofootinbib,showpacs]{revtex4}
\usepackage{amsmath, amsthm, amscd, amssymb}
\usepackage{bm}
\usepackage{bbm}

\let\vec\boldvec%

\begin{document}

\title{\bf On the Electromagnetic Properties of Matter in Collapse Models}
\author{Angelo Bassi}
\email{bassi@ts.infn.it}
\address{Department of
Physics, University of Trieste, Strada Costiera 11, 34151 Trieste, Italy \\
Istituto Nazionale di Fisica Nucleare, Trieste Section, Via Valerio 2, 34127
Trieste, Italy.}
\author{Detlef D\"urr}
\email{duerr@mathematik.uni-muenchen.de}
\address{Mathematisches Institut der L.M.U., Theresienstr. 39, 80333
M\"unchen, Germany.}
\begin{abstract}
We discuss the electromagnetic properties of both a charged free particle, and
a charged particle bounded by an harmonic potential, within collapse models. By
choosing a particularly simple---yet physically relevant---collapse model, and
under only the dipole approximation, we are able to solve the equation of
motion exactly. In this way, both the finite time and large time behavior can
be analyzed accurately. We discovered new features, which did not appear in
previous works on the same subject. Since, so far, the spontaneous photon
emission process places the strongest upper bounds on the collapse parameters,
our results call for a further analysis of this process for those atomic
systems which can be employed in experimental tests of collapse models, as well
as of quantum mechanics.
\end{abstract}
 \maketitle

\section{Introduction} \label{sec:uno}

Models of spontaneous wave function collapse~\cite{grw,www,csl,rev,pea} provide
a description of quantum (and classical) phenomena, which is free of the much
debated measurement problem affecting the standard quantum theory. This is
achieved by modifying the Schr\"odinger equation, adding nonlinear and
stochastic terms which reproduce, at a suitable scale, the process of wave
function collapse.

By modifying the Schr\"odinger equation, these models make predictions which
differ from those of standard quantum mechanics. It is interesting and
important to analyze such differences, not only for a better understanding of
these models, but also for deciding which experimental setups are more
convenient, in order to test them against quantum mechanics. Such experiments,
needless to say, would represent important tests also for the quantum theory
itself~\cite{sci}.

It has emerged from the work reported in~\cite{fu0,fu1,fu2,fu3,fu4}, that the
electromagnetic properties of matter place, so far, the strongest upper bound
on the collapse frequency $\lambda_{\text{\tiny GRW}}$ of the GRW
(Ghirardi-Rimini-Weber) model~\cite{grw} (or, equivalently, the parameter
$\gamma$ of the CSL\footnote{The GRW and CSL models are the two most popular
models of spontaneous wave function collapse. Their main difference is that the
first assumes the collapses to occur at discrete random times, through a jump
process, while the second assumes the collapse to occur continuously, through a
diffusion process.} (Continuous Spontaneous Localization) model~\cite{csl}).
More specifically, it has been proven that charged particles spontaneously emit
radiation, as a consequence of the interaction with the collapsing field, also
when according to standard quantum mechanics no radiation should be emitted;
the radiation spectrum has been computed both for a free charged
particle~\cite{fu3} and for an hydrogenic atom~\cite{fu4}. The theoretical
spectrum has been compared with available experimental data, placing an upper
bound~\cite{fu4} of only 6 orders of magnitude away from the standard CSL
value\footnote{This value for the collapse parameter $\gamma$ has been chosen
in such a way that for a single constituent---in which case the GRW and CSL
model coincide at the statistical level---the reduction occurs with the rate
$\lambda_{\text{\tiny GRW}} \simeq 2.2 \times 10^{-17} \text{s}^{-1}$ of the
GRW model~\cite{grw}. The relation between the two constants is:
$\lambda_{\text{\tiny GRW}} = \gamma (\alpha/4\pi)^{3/2}$, with $\alpha =
10^{10} \text{cm}^{-2}$~\cite{csl}. This choice implies that the two models
behave similarly, although important differences arise, due to the fact that
the GRW model refers only to systems of distinguishable particles, while the
CLS model takes into account also identical particles. The numerical value for
$\lambda_{\text{\tiny GRW}}$ has been originally chosen in such a way to ensure
that superpositions of macroscopic objects (containing roughly an Avogadro's
number of constituents) localized within the perception time of a human being,
while microscopic systems retain all their quantum properties~\cite{grw}.}
$\gamma = 10^{-30} \text{cm}^3 \text{s}^{-1}$. Note that more direct
experiments of the superposition principle of quantum mechanics, such as
diffraction experiments with macro-molecules~\cite{zei1,zei2}, place a much
weaker upper bound, which is 13 orders of magnitude away from the standard CSL
value~\cite{adler}. These figures show that analyzing the electromagnetic
properties of matter within collapse models is particularly relevant, not only
{\it per se} but also in view of future experimental tests.

The above mentioned analysis has been carried out to first order in
perturbation theory, using the CSL model. Goal of this work is to deepen our
understanding of the process of spontaneous photon emission from charged
particles. We will do it by using, in place of the CSL model, the simpler QMUPL
(Quantum Mechanics with Universal Position Localizations)
model~\cite{qmupl1,qmupl2} and we will work under the dipole approximation.
These assumptions will allows to solve the equations of motion exactly: we will
derive an exact formula for the spectrum of the emitted radiation, valid to all
orders, and we will compare it with the formulas obtained in~\cite{fu3,fu4}. As
we will see, new features will emerge, previously not discussed.

The QMUPL model of spontaneous wave function collapse applies to systems of
distinguishable non-relativistic particles. The one particle equation, which
is
sufficient for the purposes of this paper, reads:
\begin{equation} \label{eq:qmupl}
d \psi_t \; = \; \left[ - \frac{i}{\hbar} \, H \, dt +
\sqrt{\lambda} ( {\bf q} - \langle {\bf q} \rangle_t )\, d{\bf W}_t
- \frac{\lambda}{2} ( {\bf q} - \langle {\bf q} \rangle_t )^2 dt
\right] \psi_t,
\end{equation}
its generalization to a many particle system being straightforward. In the
above equation, $H$ is the standard quantum Hamiltonian of the particle, ${\bf
q}$ its position operator, ${\bf W}_t$ three independent standard Wiener
processes defined on a probability space $(\Omega, {\mathcal F}, {\mathbb P})$,
and $\lambda$ is a positive constant which sets the strength of the collapse
mechanics. The physical content of the above equation is the following. The
first term on the right-hand-side gives the usual unitary evolution, driven by
the Hamiltonian $H$. The second and third terms cause the collapse of the wave
function towards a state which is localized in {\it space}, being it driven by
the {\it position} operator. More specifically (but not entirely correctly; for
a discussion see~\cite{bd}), the third term localizes the wave function---being
it the more negative, the greater the difference $|{\bf q} - \langle{\bf
q}\rangle_t|$---while the second term, which contains the random process ${\bf
W}_t$, ensures that the collapse occurs randomly and in agreement with the Born
probability rule. The structure $({\bf q} - \langle{\bf q}\rangle_t)$ ensures
that the wave function remains normalized, even if the dynamics is not unitary
anymore.

This model, in spite of its simplicity, is particularly relevant because, in an
appropriate limit which we shall now briefly discuss, it reduces at the
statistical level to the more familiar GRW model (thus also to the CSL model,
as long as the particles are distinguishable). Here again we limit our
consideration to one single particle. The master equation describing the time
evolution of the statistical ensemble $\rho_t \equiv {\mathbb E}_{\mathbb
P}[|\psi_t\rangle \langle \psi_t|]$, where $\psi_t$ evolves according to
Eq.~\eqref{eq:qmupl}, has a Lindblad form, which in the position representation
(where $\rho_t ({\bf x},{\bf y}) \equiv \langle {\bf x} | \rho_t | {\bf y}
\rangle$) reads~\cite{qmupl2}:
\begin{equation} \label{eq:gsfdsao2}
\frac{d}{dt}\, \rho_t({\bf x},{\bf y}) \; = \; - \frac {i}{\hbar} \,
\left[ H, \rho_t({\bf x},{\bf y}) \right] \, - \, \frac{\lambda}{2}
\left({\bf x} - {\bf y} \right)^2 \rho_t({\bf x},{\bf y}).
\end{equation}
On the other hand, the one-particle GRW (and CSL) master equation
reads~\cite{grw}:
\begin{equation} \label{eq:megrw}
\frac{d}{dt} \rho_t( {\bf x}, {\bf y} ) \; = \; - \frac{i}{\hbar}
[H, \rho_t( {\bf x}, {\bf y} )] - \lambda_{\text{\tiny GRW}} \left[
1 - e^{- \frac{\alpha}{4}({\bf x} - {\bf y})^2} \right] \rho_t( {\bf
x}, {\bf y} );
\end{equation}
for the relation between the constant $\lambda_{\text{\tiny GRW}}$
characterizing the GRW model and the constant $\gamma$ defining the strength of
the collapse process in the CSL model, see the previous footnote. The second
parameter ($\alpha$) defines a correlation length $r_C = 1/\sqrt{\alpha} \simeq
10^{-5}$ cm, above which spatial superpositions are reduced.

Let us now consider situations where, for all values of ${\bf x}$ and ${\bf y}$
such that the density matrix $\rho_t({\bf x},{\bf y})$ is appreciably different
from 0, one has: $| {\bf x} - {\bf y} | \ll r_C$. We call this the {\it small
distances} assumption. This is the case if the physical system is localized
well below $r_C$, as it happens e.g. for atoms in matter. In this case, it
makes sense to take, in Eq.~\eqref{eq:megrw}, the limit $\alpha \rightarrow 0$
and $\lambda_{\text{\tiny GRW}} \rightarrow \infty$, while keeping the product
$\lambda_{\text{\tiny GRW}} \alpha$ constant. Then, Eq.~\eqref{eq:megrw}
reduces to~\eqref{eq:gsfdsao2}, with the identification:
\begin{equation}
\lambda \; = \; \frac{\alpha\lambda_{\text{\tiny GRW}}}{2} \; = \;
\frac{\alpha^{3/2}\gamma}{16 \pi^{3/2}}.
\end{equation}
Accordingly, the QMUPL model represents, at the statistical level, a good
approximation of the GRW models, for those systems which are well localized
with respect to the correlation length $r_C$.

\section{Motion of a charged particle interacting with the
electromagnetic field, bounded by a linear force, subject to
collapse in space}

In this section we explicitly solve, under only the dipole approximation, the
equations of motion for a non-relativistic charged particle interacting with
the second quantized electromagnetic field. The particle is bounded by an
harmonic potential---the limit case of a free particle will be also
discussed---and is subject to spontaneous collapses in space according to the
QMUPL model.

Eq.~\eqref{eq:qmupl} is nonlinear, but it can be appropriately reduced to a
linear (though not norm-preserving) equation trough a standard
procedure~\cite{qmupl2}. Of course, non-linearity is not canceled; it appears
back again when the statistical properties (through a change of measure) are
computed. However, since we are ultimately concerned only with physical
quantities of the type ${\mathbb E}_{\mathbb P} [ \langle \psi_t | O | \psi_t
\rangle ]$, where $O$ is any suitable self-adjoint operator, we can use the
following mathematical property. Consider the class of SDEs:
\begin{equation} \label{eq:sdez}
d \psi^{\zeta}_t = \left[ -\frac{i}{\hbar} H dt + \sqrt{\lambda}
(\zeta {\bf q} - \zeta_{\text{\tiny R}} \langle {\bf q} \rangle_t)
d{\bf W}_t - \frac{1}{2} \lambda (|\zeta|^2 {\bf q}^2 - 2 \zeta
\zeta_{\text{\tiny R}} {\bf q} \langle {\bf q} \rangle_t +
\zeta_{\text{\tiny R}}^2 \langle {\bf q} \rangle_t^2 ) dt \right]
\psi^{\zeta}_t,
\end{equation}
where $\zeta$ is a complex phase, and $\zeta_{\text{\tiny R}}$ its real part. A
straightforward application of It\^o calculus allows to prove that ${\mathbb
E}_{\mathbb P} [ \langle \psi_t | O | \psi_t \rangle ]$ is independent of
$\zeta$, in spite of the fact that Eq.~\eqref{eq:sdez} describes completely
different evolutions for the wave function, for different values of $\zeta$. In
particular, when $\zeta = 1$, Eq.~\eqref{eq:sdez} coincides with the QMUPL
collapse equation~\eqref{eq:qmupl}. On the other hand, when $\zeta = i$,
Eq.~\eqref{eq:sdez} reduces to the simpler SDE:
\begin{equation} \label{eq:ghdf}
d \psi_t = \left[ -\frac{i}{\hbar} H dt + i\sqrt{\lambda} {\bf
q}\, d{\bf W}_t - \frac{1}{2} \lambda {\bf q}^2 dt \right] \psi_t,
\end{equation}
where all the non-linear terms have disappeared. Such an equation of course
does not lead to the collapse of the wave function, since it describes a linear
and unitary\footnote{The third term of Eq.~\eqref{eq:ghdf} is an ``It\^o
term'', which disappears from the solution of the equation. For this reason the
evolution is unitary, even if apparently it does not look so.}, though
stochastic, evolution. Nevertheless it is as good as Eq.~\eqref{eq:qmupl} for
computing averages quantities. The advantage is that its linearity and
unitarity make calculations easier.

Eq.~\eqref{eq:ghdf} has to be understood in the It\^o sense. We will solve the
corresponding Stratonovich equation, where the stochastic differential $d{\bf
W}_t$ can be interpreted as the increment of a white noise ${\bf w}(t)$:
\begin{equation} \label{eq:ghdfs}
i\hbar \frac{d}{dt} \psi_t = \left[ H - \sqrt{\lambda} \hbar\,
{\bf q}\, {\bf w}(t)  \right] \psi_t.
\end{equation}
This is a standard Schr\"odinger equation with a random potential depending on
the position ${\bf q}$ of the particle. Note that the last term of
Eq.~\eqref{eq:ghdf} has disappeared in going from the It\^o to the
Stratonovich
formulation of the SDE. Actually, one can be more general and assume that
${\bf
w}(t)$ represent three Gaussian noises with zero mean and a general
correlation
function, without having to change the mathematical formalism. However, this
goes beyond the scope of the present analysis, so we will keep assuming that
${\bf w}(t)$ are white noises.

Coming back to our physical system, the standard Hamiltonian $H$ is:
\begin{equation} \label{eq:H}
H \; = \; \frac{1}{2m_0}({\bf p} - e {\bf A})^2 + \frac{1}{2}
\kappa\, {\bf q}^2 + \frac{1}{2}\epsilon_0 \int d^3x \, \left[ {\bf
E}^2 + c^2 {\bf B}^2 \right],
\end{equation}
where $m_0$ is the bare mass of the particle, $\kappa$ is the force constant of
the harmonic term, ${\bf A}$ is the vector potential, ${\bf E}$ and ${\bf B}$
the electric and magnetic fields respectively, $e$ is the electric charge, $c$
the speed of light and $\epsilon_0$ the vacuum permittivity. Throughout this
section, we use the gauge: ${\bf \nabla} \cdot {\bf A} = 0$ and $V = 0$, where
$V$ is the electromagnetic scalar potential\footnote{Note that we are assuming
that the spontaneous collapse process occurs only for the particle, not for the
electromagnetic field. The reason is that, so far, collapse models have been
considered only for massive particles, being the localization of their wave
function sufficient for solving the measurement problem of quantum mechanics.
However, in a more speculative scenario, e.g. where the collapse mechanism is
linked to gravitational phenomena, one could assume that also the photons' wave
function undergoes a spontaneous localization process.}.

The plane wave decomposition of the vector potential ${\bf A}$
reads:
\begin{equation} \label{eq:A}
{\bf A}({\bf x}) \; = \; \sqrt{\frac{\hbar}{\epsilon_0}} \sum_{\mu}
\int \frac{d^3k}{\sqrt{(2\pi)^3}}\, \frac{1}{\sqrt{2\omega_k}}\,
\vec{\epsilon}_{\bf k \mu} \left[ a_{\bf k \mu}^{\phantom \dagger}
e^{i {\bf k} \cdot {\bf x}} + a_{\bf k \mu}^{\dagger} e^{-i {\bf k}
\cdot {\bf x}} \right],
\end{equation}
where $\omega_k = c k$ ($k = | {\bf k} |$) is the frequency
corresponding to the wave vector ${\bf k}$; $\vec{\epsilon}_{\bf k
\mu}$ ($\mu = 1,2$) are the linear polarization vectors and $a_{\bf
k \mu}^{\dagger}$, $a_{\bf k \mu}^{\phantom \dagger}$ are the
creation and annihilation operators satisfying the standard
commutation relations:
\begin{equation}
[a_{\bf k \mu}^{\phantom \dagger}, a_{\bf k' \mu'}^{\dagger}] \; =
\; \delta_{\mu \mu'}^{\phantom \dagger} \delta^{(3)}({\bf k} - {\bf
k}').
\end{equation}

Up to now the model is exact, but not exactly solvable. To further proceed in
the analysis, we make the {\it dipole approximation} $e^{i {\bf k} \cdot {\bf
x}} \simeq 1$, which holds as long as the wave-length of the electromagnetic
radiation is much larger than the typical size of an atom. Note that this
assumption is compatible with the {\it small distances} assumption discussed in
the previous section. The resulting model turns out to be ultraviolet
divergent: we cure this problem by introducing a form factor $g({\bf k})$,
corresponding to the Fourier transform of the charge distribution (normalized
to unity):
\begin{equation}
g({\bf k}) \; := \; \int \frac{d^3 k}{\sqrt{(2\pi)^3}} \, \rho({\bf
r}) \, e^{-i {\bf k} \cdot {\bf r}}, \qquad \int d^3 r\, \rho({\bf r}) = 1.
\end{equation}
Under these approximations, the vector potential~\eqref{eq:A}
becomes:
\begin{equation}
{\bf A}({\bf x}) \; = \; \sqrt{\frac{\hbar}{\epsilon_0}} \sum_{\mu}
\int d^3k\, \frac{g({\bf k})}{\sqrt{2\omega_k}}\,
\vec{\epsilon}_{\bf k \mu} \left[ a_{\bf k \mu}^{\phantom \dagger} +
a_{\bf k \mu}^{\dagger} \right];
\end{equation}
and the factor $(2\pi)^{-3/2}$ as been included in the definition of
$g({\bf k})$. In this way, in the point-particle limit, $g({\bf k})
\rightarrow 1/\sqrt{(2\pi)^3}$.

Since the total Hamiltonian in Eq.~\eqref{eq:ghdfs} is a standard---though
stochastic---Hamiltonian, one can conveniently work in the Heisenberg picture.
The equations of motions for the position ${\bf q}(t)$ of the particle and of
the conjugate momentum ${\bf p}(t)$ can be immediately derived:
\begin{eqnarray}
\frac{d {\bf p}}{dt} & = & - \kappa\, {\bf q} \; + \; \sqrt{\lambda}
\hbar {\bf w}(t), \label{eq:dfgsd}\\
& & \nonumber \\ \frac{d {\bf q}}{dt} & = & \frac{\bf p}{m_0} \; - \;
\frac{e}{m_0}\, {\bf A}, \label{eq:dfgsd2}
\end{eqnarray}
while the equation of motion for the electromagnetic-field operator
$a_{\bf k \mu}^{\dagger}(t)$ is:
\begin{equation} \label{eq:dfgsd3}
\frac{d a_{\bf k \mu}^{\dagger}}{dt} = {\phantom -}i \omega_{\bf k}
a_{\bf k \mu}^{\dagger} - \frac{ie}{\sqrt{\hbar\epsilon_0} m_0}\,
\frac{g({\bf k})}{\sqrt{2\omega_k}} \vec{\epsilon}_{\bf k
\mu}^{\phantom \dagger} \cdot {\bf p} \; + \; \frac{i
e^2}{\epsilon_0 m_0} \, \frac{g({\bf k})}{\sqrt{2\omega_k}}
\vec{\epsilon}_{\bf k \mu}^{\phantom \dagger} \cdot \sum_{\mu'} \int
d^3 k' \frac{g({\bf k}')}{\sqrt{2\omega_{k'}}} \vec{\epsilon}_{\bf
k' \mu'}^{\phantom \dagger} \left[ a_{\bf k' \mu'}^{\phantom
\dagger} + a_{\bf k' \mu'}^{\dagger} \right];\;\;\;\;
\end{equation}
the equation for $a_{\bf k \mu}^{\phantom \dagger}(t)$ can be
obtained from the previous one by taking the hermitian conjugate.
The above set of coupled linear differential equations can be
conveniently solved with the help of the Laplace transform; the
equations for the transformed variables (which are denoted by a
tilde) read:
\begin{eqnarray}
z \tilde{\bf p}(z) - {\bf p}(0) & = & - \kappa \tilde{\bf q}(z) +
\hbar \sqrt{\lambda} \tilde{\bf w}(z), \\ & & \nonumber \\
z \tilde{\bf q}(z) - {\bf q}(0) & = & \frac{\tilde{\bf p}(z)}{m_0} -
\frac{e}{m_0} \sqrt{\frac{\hbar}{\epsilon_0}} \sum_{\mu} \int d^3k
\frac{1}{\sqrt{2 \omega_k}}\, g({\bf k})\, \vec{\epsilon}_{{\bf
k}\mu} \left[ \tilde{a}^{\dagger}_{{\bf k}\mu}(z) +
\tilde{a}^{\phantom \dagger}_{{\bf k}\mu}(z) \right], \\ & & \nonumber \\
z \tilde{a}^{\dagger}_{{\bf k}\mu}(z) - {a^{\dagger}_{{\bf
k}\mu}}(0) & = &  i \omega_k \, \tilde{a}^{\dagger}_{{\bf k}\mu}(z)
- \frac{ie}{\sqrt{\hbar \epsilon_0}}\, \frac{1}{\sqrt{2 \omega_k}}
\, g({\bf k}) \, \vec{\epsilon}_{{\bf k} \mu}\cdot \left[ z
\tilde{\bf q}(z) - {\bf q}(0) \right], \\ & & \nonumber \\
z \tilde{a}^{\phantom \dagger}_{{\bf k}\mu}(z) - {a^{\phantom
\dagger}_{{\bf k}\mu}}(0) & = & - i \omega_k \, \tilde{a}^{\phantom
\dagger}_{{\bf k}\mu}(z) + \frac{ie}{\sqrt{\hbar \epsilon_0}}\,
\frac{1}{\sqrt{2 \omega_k}} \, g({\bf k}) \, \vec{\epsilon}_{{\bf k}
\mu}\cdot \left[ z \tilde{\bf q}(z) - {\bf q}(0) \right],
\end{eqnarray}
where $z$ is the transformed time. The above set now represents a
system of coupled algebraic equations, which can be solved in a
standard way. The calculation is long but straightforward;
transforming back to the original variables, one obtains:
\begin{eqnarray}
{\bf q}(t) & = & \left[ 1 - \kappa\, F_1(t) \right] {\bf q}(0) \;
+ \; F_0(t)\, {\bf p}(0) \nonumber \\
& & - e\, \sqrt{\frac{\hbar}{\epsilon_0}} \, \sum_{\mu} \int d^3 k
\, \frac{g({\bf k})}{\sqrt{2\omega_k}} \, \vec{\epsilon}_{\bf k
\mu}^{\phantom \dagger} \left[ G^{+}_{1}(k,t)\, a_{\bf k
\mu}^{\phantom \dagger}(0) +
G^{-}_{1}(k,t)\, a_{\bf k \mu}^{\dagger}(0) \right] \nonumber \\
& & + \sqrt{\lambda}\hbar \int_0^t ds F_0(t - s) {\bf
w}(s), \label{eq:fsdgh} \\ & & \nonumber\\
{\bf p}(t) & = & - \kappa \left[ t - \kappa\, F_2(t) \right] {\bf
q}(0) \; + \; \left[ 1 - \kappa\, F_1(t) \right] {\bf p}(0)
\nonumber \\
& & + \kappa\, e \, \sqrt{\frac{\hbar}{\epsilon_0}} \, \sum_{\mu}
\int d^3 k \, \frac{g({\bf k})}{\sqrt{2\omega_k}} \,
\vec{\epsilon}_{\bf k \mu}^{\phantom \dagger} \left[
G^{+}_{0}(k,t)\, a_{\bf k \mu}^{\phantom \dagger}(0) +
G^{-}_{0}(k,t)\, a_{\bf k \mu}^{\dagger}(0) \right] \nonumber \\
& & + \sqrt{\lambda} \hbar \int_0^t ds \left[ 1 - \kappa\,
F_1(t-s) \right] {\bf w}(s) \\ & & \nonumber \\
a_{\bf k \mu}^{\dagger}(t) & = & e^{i \omega_k t} a_{\bf k
\mu}^{\dagger}(0) \; - \; \frac{ie}{\sqrt{\hbar\epsilon_0}}\,
\frac{g({\bf k})}{\sqrt{2\omega_k}} \, \vec{\epsilon}_{\bf k
\mu}^{\phantom \dagger} \cdot \left[ G^{-}_{1}(k,t)\, {\bf p}(0) -
\kappa \, G^{-}_{0}(k,t)
\, {\bf q}(0) \right] \nonumber \\
& & + \frac{i e^2}{\epsilon_0} \, \frac{g({\bf
k})}{\sqrt{2\omega_k}} \, \vec{\epsilon}_{\bf k \mu}^{\phantom
\dagger} \cdot \sum_{\mu'} \int d^3 k' \, \frac{g({\bf
k}')}{\sqrt{2\omega_{k'}}} \, \vec{\epsilon}_{\bf k' \mu'}^{\phantom
\dagger} \left[ G^{-}_{+}(k,k',t)\, a_{\bf k' \mu'}^{\phantom
\dagger}(0) +
G^{-}_{-}(k,k',t)\, a_{\bf k' \mu'}^{\dagger}(0) \right] \nonumber \\
& & - ie\sqrt{\frac{\hbar \lambda}{\epsilon_0}} \, \frac{g({\bf
k})}{\sqrt{2\omega_k}} \, \vec{\epsilon}_{\bf k \mu}^{\phantom
\dagger} \cdot \int_0^t ds\,
G^{-}_{1}(k,t-s) {\bf w}(s), \label{eq:fed} \\ & & \nonumber \\
a_{\bf k \mu}^{\phantom \dagger}(t) & = & e^{-i \omega_k t} a_{\bf k
\mu}^{\phantom \dagger}(0) \; + \;
\frac{ie}{\sqrt{\hbar\epsilon_0}}\, \frac{g({\bf
k})}{\sqrt{2\omega_k}} \, \vec{\epsilon}_{\bf k \mu}^{\phantom
\dagger} \cdot \left[ G^{+}_{1}(k,t)\, {\bf p}(0) - \kappa \,
G^{+}_{0}(k,t)
\, {\bf q}(0) \right] \nonumber \\
& & - \frac{i e^2}{\epsilon_0} \, \frac{g({\bf
k})}{\sqrt{2\omega_k}} \, \vec{\epsilon}_{\bf k \mu}^{\phantom
\dagger} \cdot \sum_{\mu'} \int d^3 k' \, \frac{g({\bf
k}')}{\sqrt{2\omega_{k'}}} \, \vec{\epsilon}_{\bf k' \mu'}^{\phantom
\dagger} \left[ G^{+}_{+}(k,k',t)\, a_{\bf k' \mu'}^{\phantom
\dagger}(0) +
G^{+}_{-}(k,k',t)\, a_{\bf k' \mu'}^{\dagger}(0) \right] \nonumber \\
& & + ie\sqrt{\frac{\hbar \lambda}{\epsilon_0}} \, \frac{g({\bf
k})}{\sqrt{2\omega_k}} \, \vec{\epsilon}_{\bf k \mu}^{\phantom
\dagger} \cdot \int_0^t ds\, G^{+}_{1}(k,t-s) {\bf w}(s),
\label{eq:fe}
\end{eqnarray}
In the previous formulas, we have introduced the following
functions:
\begin{eqnarray}
F_n(t) & = & \int_{\Gamma} \frac{dz}{2\pi i} \, \frac{e^{zt}}{z^n
H(z)}, \qquad\qquad\qquad\qquad\;\;\; n = 0,1,2, \label{eq:F}\\
G^{\pm}_n(k,t) & = & \int_{\Gamma} \frac{dz}{2\pi i} \, \frac{z^n
e^{zt}}{(z \pm i \omega_k) H(z)}, \qquad\qquad\qquad n = 0,1, \label{eq:G} \\
G^{\pm}_{\pm}(k,k't) & = & \int_{\Gamma} \frac{dz}{2\pi i} \,
\frac{z^2 e^{zt}}{(z \pm i \omega_k)(z \pm i \omega_{k'}) H(z)};
\label{eq:Gpm}
\end{eqnarray}
in the third expression, the upper $\pm$ refers to the first parenthesis, while
the lower one refers to the second parenthesis. In all the above formulas,
according to the theory of Laplace transform, the contour $\Gamma$ must be a
line parallel to the imaginary axis, lying to the right of all singularities of
the integrand. The above solutions should be compared with those obtained
in~\cite{pol1,pol2}, where the collapse process was not taken into account: a
part from a marginal calculational mistake in~\cite{pol2} in the evolution of
${\bf p}(t)$, the two results agree when $\lambda$ is set to zero in
Eqs.~\eqref{eq:fsdgh}--~\eqref{eq:fe}.

The function $H(z)$ is defined as follows:
\begin{equation} \label{eq:dfxs}
H(z) \; = \; \kappa + z^2 \left[ m_0 + \frac{8\pi e^2}{3\epsilon_0}
\int_0^{\infty} dk\, g(k)^2 \, \frac{k^2}{z^2 + \omega_k^2} \right]
\end{equation}
(from now on we assume the form factor to depend only on the modulus $k$ of
${\bf k}$). This is a crucial quantity, as through
formulas~\eqref{eq:F}--\eqref{eq:Gpm} it determines the time evolution of all
physical quantities. It depends on the form factor $g(k)$: simply removing it,
would make the integral ultraviolet divergent. To overcome the problem, we
apply a renormalization procedure. The quantity within square brackets
in~\eqref{eq:dfxs} can be rewritten as follows:
\begin{equation} \label{eq:dfgsdfg2}
m_0 + \frac{8\pi e^2}{3\epsilon_0} \int_0^{\infty} dk\, g(k)^2 \,
\frac{k^2}{z^2 + \omega_k^2} =  \left( m_0 + \frac{4}{3}m_r \right) -
\frac{8\pi e^2}{3\epsilon_0 c^2}\, z^2 \int_0^{\infty} dk\,
\frac{g(k)^2}{z^2 + \omega_k^2}.
\end{equation}
where $m_r$ is the electrostatic mass:
\begin{equation}
m_r \; := \; \frac{e^2}{8\pi\epsilon_0c^2} \int d^3r d^3r' \,
\frac{\rho({\bf r})\rho({\bf r}')}{|{\bf r} - {\bf r}'|} \; \equiv
\; \frac{2\pi e^2}{\epsilon_0 c^2} \int d^3k \, g(k)^2.
\end{equation}
When  $g({\bf k}) \rightarrow 1/\sqrt{(2\pi)^3}$, $m_r$ diverges. We apply the
classical renormalization procedure\footnote{One can notice that the collapse
terms do not enter the following equations, thus the renormalization procedure
applies like in standard cases.} for a non-relativistic charged particle
coupled to the electromagnetic field~\cite{roh,spo,mil} (which is valid both as
a classical calculation and as a Heisenberg picture, quantum mechanical one,
like in our case). According to it, as $m_r \rightarrow + \infty$ in the
point-particle limit, one assumes that $m_0 \rightarrow - \infty$, in such a
way that $m := m_0 + (4/3)m_r$ remains finite. This is assumed to be the
renormalized mass.

The last term of~\eqref{eq:dfgsdfg2} instead remains finite in the limit, the
integral can be evaluated, and $H(z)$ takes the well-behaved expression:
\begin{equation} \label{eq:Hfun}
H(z) \; = \; \kappa + z^2 \left[ m - \beta \, z \right],
\qquad\qquad \beta \; = \; \frac{e^2}{6 \pi \epsilon_0 c^3} \;
\simeq \; 5.71 \times 10^{-54}\, \makebox{Kg s}.
\end{equation}
Note that $\beta$ is precisely the coefficient in front of the Abraham-Lorentz
force, which is responsible for the runaway behavior of the corresponding
Abraham-Lorentz equation, as we shall soon see. $H(z)$ is a polynomial of third
degree, whose zeros can be found by the standard Cardan method. One solution is
real and two are complex conjugate. Let $\omega_0 := \sqrt{\kappa/m}$ be the
frequency of the oscillator. By assuming $\omega_0 \ll 2 m/ \sqrt{27} \beta
\simeq 6.14 \times 10^{22}$ s$^{-1}$ for an electron ($\hbar \omega_0 \ll 4.04
\times 10^4$ KeV), their approximate value is (see Appendix A):
\begin{equation} \label{eq:sol}
z_1 \; \simeq \; \frac{m}{\beta} \; + \;  o(\omega_0), \quad \qquad
z_{2,3} \; \simeq \; - \frac{\omega_0^2 \beta}{2m} \, \pm \, i
\omega_0 \; + \; o(\omega_0^3)
\end{equation}
Given the above results, the functions $F_n(t)$ and $G^{\pm}_n(k,t)$
defined in~\eqref{eq:F} and~\eqref{eq:G}, which are the only ones we
will use in the subsequent analysis, become:
\begin{eqnarray}
F_n(t) & = &  \sum_{\ell = 1}^3 z_{\ell} e^{z_{\ell}t} \left[
\frac{z - z_{\ell}}{H(z)} \right]_{z = z_{\ell}} + \; \left\{
\begin{array}{cl}
0 & n = 0, \\
\omega_0^{-2} & n = 1, \\
t \omega_0^{-2} & n = 2,
\end{array}
\right. \label{eq:Ff}\\ & & \nonumber \\
G^{\pm}_n(k,t) & = & \sum_{\ell = 1}^3 \frac{z_{\ell}^n
e^{z_{\ell}t}}{(z_{\ell} \pm i k)} \left[ \frac{z - z_{\ell}}{H(z)}
\right]_{z = z_{\ell}} + \; \frac{(\pm i k)^n e^{\pm i k t}}{H(\pm i
k)}. \label{eq:Gf}
\end{eqnarray}
The term in~\eqref{eq:Ff} and~\eqref{eq:Gf} with $\ell = 1$ diverges
exponentially, since $z_1$ is positive. As we have anticipated, this is a
manifestation of the runaway behavior of the Abraham-Lorentz
equation~\cite{roh,spo,mil,jac}. In particular, in the free particle case
($\omega_0 = 0 $), the coefficient $z_1$ corresponds to the rate of exponential
growth of the acceleration, as discussed in textbooks. This problem is still
open, and we pragmatically dismiss it by ignoring, in the subsequent formulas,
all terms with $\ell = 1$.

\subsection{The spectrum of the spontaneously emitted radiation}

We are now in a position to compute the spectrum of the radiation spontaneously
emitted by the particle, due to the interaction with the noise. Let $N_{\bf k
\mu}^{\phantom \dagger}(t) := a_{\bf k \mu}^{\dagger}(t) a_{\bf k
\mu}^{\phantom \dagger}(t)$ be the density of photons of wave vector ${\bf k}$
and polarization $\mu$. Let $|\phi\rangle := |\psi_{\text{\tiny ho}}\rangle
|\Omega\rangle$ be the initial state of the system, where $|\psi_{\text{\tiny
ho}}\rangle$ is the initial state of the harmonic oscillator and
$|\Omega\rangle$ is the vacuum state for the electromagnetic field. Let finally
$S({\bf k}, \mu, t) := {\mathbb E}_{\mathbb P} [ \langle \phi | N_{\bf k
\mu}^{\phantom \dagger}(t) | \phi \rangle ]$, be the spectrum of the emitted
radiation, averaged over the noise. By inspecting Eqs.~\eqref{eq:fed}
and~\eqref{eq:fe}, one can notice that all terms of $N_{\bf k \mu}^{\phantom
\dagger}(t)$ containing either $a_{\bf k \mu}^{ \dagger}(0)$ or $a_{\bf k
\mu}^{\phantom \dagger}(0)$ give a zero contribution, when averaged with
respect to the vacuum state, while all terms containing ${\bf w}(t)$ give a
zero contribution, when the stochastic average is taken. Accordingly, $S({\bf
k}, \mu, t)$ is the sum of two terms:
\begin{equation}
S({\bf k}, \mu, t) \; = \; S_{\text{\tiny qm}}({\bf k}, \mu, t) \, +
\, S_{\text{\tiny col}}({\bf k}, \mu, t),
\end{equation}
where $S_{\text{\tiny qm}}({\bf k}, \mu, t)$ is the standard quantum
formula, while $S_{\text{\tiny col}}({\bf k}, \mu, t)$ is the
contribution due to the noise. We are interested in computing this
second term, which reads:
\begin{equation} \label{eq:ser}
S_{\text{\tiny col}}({\bf k}, \mu, t) \; = \; \frac{\lambda \hbar
e^2}{16 \pi^3\epsilon_0}\, \frac{1}{\omega_k}\, \int_0^t ds
G^{-}_{1}(k,t-s) G^{+}_{1}(k,t-s),
\end{equation}
This is the main formula. In the next section, we will apply it to
the two interesting cases of a free particle ($\omega_0 = 0)$ and of
a bounded particle ($\omega_0 \neq 0)$.

\section{The free particle}

The free particle evolution can be deduced from the previous
formulas by taking the limit $\omega_0 \rightarrow 0$. However it
turns out to be easier to re-do the calculation, starting from
Eqs.~\eqref{eq:dfgsd}--\eqref{eq:dfgsd3} with $\kappa = 0$. The
final result is:
\begin{eqnarray}
{\bf q}(t) & = &  {\bf q}(0) + \bar{F}_0(t) {\bf p}(0) - e
\sqrt{\frac{\hbar}{\epsilon_0}} \, \sum_{\mu} \int d^3 k \,
\frac{g({\bf k})}{\sqrt{2\omega_k}} \, \vec{\epsilon}_{\bf k
\mu}^{\phantom \dagger} \left[ \bar{G}^{+}_{1}(k,t)\, a_{\bf k
\mu}^{\phantom \dagger}(0) + \bar{G}^{-}_{1}(k,t)\, a_{\bf k
\mu}^{\dagger}(0) \right] \nonumber \\
& & + \sqrt{\lambda}\hbar \int_0^t ds \bar{F}_0(t - s) {\bf
w}(s), \label{eq:fdsdzfa}\\ & & \nonumber \\
{\bf p}(t) & = &  {\bf p}(0) + \sqrt{\lambda}\hbar \int_0^t ds\,
{\bf w}(s), \\  & & \nonumber \\
a_{\bf k \mu}^{\dagger}(t) & = & e^{i \omega_k t} a_{\bf k
\mu}^{\dagger}(0) \; - \; \frac{ie}{\sqrt{\hbar\epsilon_0}}\,
\frac{g({\bf k})}{\sqrt{2\omega_k}} \, \bar{G}^{-}_{1}(k,t)\,
\vec{\epsilon}_{\bf k \mu}^{\phantom \dagger} \cdot {\bf
p}(0)  \nonumber \\
& & + \frac{i e^2}{\epsilon_0} \, \frac{g({\bf
k})}{\sqrt{2\omega_k}} \, \vec{\epsilon}_{\bf k \mu}^{\phantom
\dagger} \cdot \sum_{\mu'} \int d^3 k' \, \frac{g({\bf
k}')}{\sqrt{2\omega_{k'}}} \, \vec{\epsilon}_{\bf k' \mu'}^{\phantom
\dagger} \left[ \bar{G}^{-}_{+}(k,k',t)\, a_{\bf k' \mu'}^{\phantom
\dagger}(0) +
\bar{G}^{-}_{-}(k,k',t)\, a_{\bf k' \mu'}^{\dagger}(0) \right] \nonumber \\
& & - ie\sqrt{\frac{\hbar \lambda}{\epsilon_0}} \, \frac{g({\bf
k})}{\sqrt{2\omega_k}} \, \vec{\epsilon}_{\bf k \mu}^{\phantom
\dagger} \cdot \int_0^t ds\,
\bar{G}^{-}_{1}(k,t-s) {\bf w}(s), \label{eq:fedfr} \\ & & \nonumber \\
a_{\bf k \mu}^{\phantom \dagger}(t) & = & e^{-i \omega_k t} a_{\bf k
\mu}^{\phantom \dagger}(0) \; + \;
\frac{ie}{\sqrt{\hbar\epsilon_0}}\, \frac{g({\bf
k})}{\sqrt{2\omega_k}} \, \bar{G}^{+}_{1}(k,t)\, \vec{\epsilon}_{\bf
k \mu}^{\phantom
\dagger} \cdot {\bf p}(0) \nonumber \\
& & - \frac{i e^2}{\epsilon_0} \, \frac{g({\bf
k})}{\sqrt{2\omega_k}} \, \vec{\epsilon}_{\bf k \mu}^{\phantom
\dagger} \cdot \sum_{\mu'} \int d^3 k' \, \frac{g({\bf
k}')}{\sqrt{2\omega_{k'}}} \, \vec{\epsilon}_{\bf k' \mu'}^{\phantom
\dagger} \left[ \bar{G}^{+}_{+}(k,k',t)\, a_{\bf k' \mu'}^{\phantom
\dagger}(0) +
\bar{G}^{+}_{-}(k,k',t)\, a_{\bf k' \mu'}^{\dagger}(0) \right] \nonumber \\
& & + ie\sqrt{\frac{\hbar \lambda}{\epsilon_0}} \, \frac{g({\bf
k})}{\sqrt{2\omega_k}} \, \vec{\epsilon}_{\bf k \mu}^{\phantom
\dagger} \cdot \int_0^t ds\, \bar{G}^{+}_{1}(k,t-s) {\bf w}(s),
\label{eq:fefr}
\end{eqnarray}
with:
\begin{eqnarray}
\bar{F}_0(t) & = & \frac{t}{m} + \frac{\beta^2}{m^2} e^{mt/\beta},
\label{eq:fdsfds1}
\\
\bar{G}^{\pm}_{1}(k,t) & = & \mp \frac{i}{m \omega_{\kappa}} \mp
\frac{i e^{\mp i \omega_{\kappa} t}}{\omega_{\kappa} (m \pm i
\omega_{\kappa})} + \frac{e^{mt/\beta}}{(m/\beta)[(m/\beta)\pm i
\omega_{\kappa}]}, \label{eq:fdsfds2} \\
\bar{G}^{\pm}_{\pm}(k,k',t) & = & \frac{e^{\mp i \omega_{\kappa}
t}}{i (\mp \omega_{\kappa} \pm \omega_{\kappa'})(m \pm i \beta
\omega_{\kappa})} + \frac{e^{\mp i \omega_{\kappa'} t}}{i (\mp
\omega_{\kappa'} \pm \omega_{\kappa})(m \pm i \beta
\omega_{\kappa'})} \nonumber \\
& & + \frac{e^{mt/\beta}}{[(m/\beta)\pm i \omega_{\kappa}][(m/\beta)
\pm i \omega_{\kappa'}]}. \label{eq:fdsfds3}
\end{eqnarray}
In the last expression, the upper $\pm$ refers to the sign in front
of each $\omega_{\kappa}$, while the lower $\pm$ refers to the sign
in front of each $\omega_{\kappa'}$. Once again, in all above
formulas we have a run-away behavior, as consequence of the
renormalization procedure. In the subsequent analysis, we neglect
such terms.

There are two quantities which are of particular interest, in order
to understand the behavior of the free charged particle under the
influence of the collapsing field: the evolution of the mean kinetic
energy, and the spectrum of the emitted radiation. We shall now
discuss both of them.

\subsection{The mean free kinetic energy}

The mean kinetic energy of the particle is given by:
\begin{equation}
E_{\text{\tiny mean}}(t) \; \equiv \; \frac{1}{2}\, m\, {\mathbb
E}[\langle \phi|{\bf \dot{q}}(t)^2| \phi \rangle].
\end{equation}
From Eqs.~\eqref{eq:fdsdzfa} and~\eqref{eq:fdsfds1} we have:
\begin{eqnarray}
{\bf \dot{q}}(t) & = & \frac{{\bf p}(0)}{m} - e
\sqrt{\frac{\hbar}{\epsilon_0}} \, \sum_{\mu} \int d^3 k \,
\frac{g({\bf k})}{\sqrt{2\omega_k}} \, \vec{\epsilon}_{\bf k
\mu}^{\phantom \dagger} \left[ \frac{e^{-i\omega_k
t}}{m+i\beta\omega_k}\, a_{\bf k \mu}^{\phantom \dagger}(0) +
\frac{e^{i\omega_k t}}{m-i\beta\omega_k}\, a_{\bf k
\mu}^{\dagger}(0) \right] \nonumber \\
& & + \frac{\sqrt{\lambda}\hbar}{m} \int_0^t ds \, {\bf w}(s).
\end{eqnarray}
By taking as initial state $|\phi\rangle = |\psi_{\text{\tiny free}}\rangle
|\Omega\rangle$, as in the previous section, and after differentiating over
time, one obtains the following expression:
\begin{equation}
\frac{d}{dt} E_{\text{\tiny mean}} \; = \; \frac{3}{2}\frac{\lambda
\hbar^2}{m} \; = \; \frac{3}{4}\frac{\lambda_{\text{\tiny GRW}}
\alpha_{\text{\tiny GRW}} \hbar^2}{m},
\end{equation}
which corresponds to the standard GRW formula~\cite{grw}. We have then a very
interesting result: in spite the fact that---as we shall see in the next
subsection---the particle emits radiation at a constant rate, its mean kinetic
energy increases steadily in time as if the particle were neutral. In other
words, the noise drives enough energy into the particle both to increase its
kinetic energy and to make it radiate. This is a consequence of the fact that
the collapse terms contain only the position operator ${\bf q}$, due to which
${\bf w}_t$ acts like an {\it infinite temperature} noise; this feature has
been first pointed out in~\cite{bvi}. In the same reference, it was shown that
a term proportional to the momentum operator acts like a dissipative term,
thanks to which the mean energy thermalizes to a finite value, associated to a
temperature which can be considered as the temperature of the noise. This is
similar to what happens in the theory of quantum Brownian
motion~\cite{bv1,bv2,bv3}, and more generally in the theory open quantum
systems, which does not come as a surprise, since collapse models and open
quantum systems rely on similar master equations.

The above results can be read in two different way. On a more
conservative level, one can accept this steady energy increase as a
feature of the model, as long as it does not violate known
experimental data. On a more speculative level, it suggests that the
coupling between the noise and the wave function should be modified
in order for the total energy (energy of the noise, plus kinetic
energy of the particle, plus energy of the emitted radiation) to be
conserved. According to this view, the models so far proposed (GRW,
CSL, QMUPL) are first approximations of more realistic models of
spontaneous wave function collapse, yet to be formulated.

\subsection{The spectrum of the emitted radiation}

By using Eq.~\eqref{eq:ser}, with $G^{\pm}_{1}(k,t)$ given by
Eq.~\eqref{eq:fdsfds2}, we obtain the following expression for the
time derivative of the emitted spectrum:
\begin{eqnarray}
\frac{d}{dt}S_{\text{\tiny col}}({\bf k}, \mu, t) & = &
\frac{\lambda \hbar e^2}{16 \pi^3\epsilon_0}\, \frac{1}{\omega_k}\,
\left[ \frac{2m^2 + \beta^2 \omega_k^2}{m^2 \omega_k^2 (m^2 +
\beta^2 \omega_k^2)} + \frac{2\beta}{m \omega_k^2 (m^2 + \beta^2
\omega_k^2)} \,
\omega_k \sin \omega_k t \right. \nonumber \\
& & \left. - \frac{2}{\omega_k^3 (m^2 + \beta^2 \omega_k^2)} \,
\omega_k \cos \omega_k t \right].
\end{eqnarray}
Since all observations are made over a period of time~\cite{fu3} much longer
than the characteristic photon's frequencies, the two oscillating terms in the
above expression average to 0. We are then left with only the first expression
within brackets.

The physically interesting quantity is the spontaneous
photon-emission rate $d \Gamma_k / dk$ per unit photon momentum.
This is obtained from $dS_{\text{\tiny col}}({\bf k}, \mu, t)/dt$ by
summing over the polarization states and integrating over all
directions in the photon's momentum space. The final result is:
\begin{equation} \label{eq:wnfr}
\frac{d}{dk}\, \Gamma_k \; = \; \frac{\lambda \hbar e^2}{2 \pi^2
\epsilon_0 m^2 c^3 k}\, \cdot \, \frac{2 + (\beta c k / m)^2}{1 +
(\beta c k / m)^2}.
\end{equation}
It reassembles Eq.~(21) of~\cite{fu2} (and Eq. (3.14) of~\cite{fu1}), when
replacing $\epsilon_0 \rightarrow 1/4\pi$ because of the different system of
units used, and when taking $\lambda = (m/m_{\text{\tiny N}})^2 \lambda_0$
($m_{\text{\tiny N}}$ is the nucleon mass) as assumed in the mass-dependent CSL
model~\cite{fu0}. The only difference is the extra factor $[2 + (\beta c k /
m)^2]/[1 + (\beta c k/m)^2]$, the $\beta $ dependence in which comes about
because the result of~\cite{fu1,fu2} has been carried out only to first
perturbative order, while our result is exact (within the dipole
approximation). For an electron, $ (\beta c k/m)^2 \; \simeq \; (9.47 \times
10^{-6} E_k/\text{KeV})^2$, where $E_k = \hbar c k$ is the energy of a photon
of momentum $k$. Table I of~\cite{fu1}  reports data from photons in an energy
range between 11 and 501 KeV: our calculation shows that, in this range, the
first-order perturbation theory is extremely accurate.

Since Eq.~\eqref{eq:wnfr} is valid for finite times, it provides a trustable
understanding of the radiation process within the limits of the dipole
approximation, i.e. as long as the particle does not move too fast, or as long
as the photon's  momentum is not too large. By keeping only the leading terms
in the relevant parameters, i.e. by setting $\beta = 0$, Eq.~\eqref{eq:wnfr}
reduces to twice the large--time, first--order CSL expression of~\cite{fu3}
and~\cite{fu4}. However, according to the argument of Sec.~\ref{sec:uno}, the
CSL and QMUPL models should agree for sufficiently well localized
systems\footnote{ One can argue that the free particle case contradicts this
assumption, as the wave function of a free particle rapidly spreads out in
space; however, at least for sufficiently short times the approximation is
correct.} (with respect to the scale set by $r_C \simeq 10^{-5}\text{cm}$); the
origin of this discrepancy will be the subject of further exploration.

As a last comment, we note that Eq.~\eqref{eq:wnfr} predicts an infinite amount
of energy to be emitted per unit time, as $d \Gamma_k / dk$ is of order $1/k$
for large $k$. This ultraviolet catastrophe is a consequence of the dipole
approximation. One of the effects of the term $e^{i {\bf k} \cdot {\bf x}}$ in
Eq.~\eqref{eq:A} is to temper the electromagnetic coupling for high
frequencies; by replacing $e^{i {\bf k} \cdot {\bf x}}$ with 1, this effect is
neglected. Accordingly, Eq.~\eqref{eq:wnfr} is not trustable anymore in the
very large $k$ limit.

\section{The harmonic oscillator}

When the particle is bounded by a linear force, the emitted spectrum
takes a quite different expression. By inserting Eqs.~\eqref{eq:Gf},
and ignoring the term $\ell = 1$ which gives a runaway solution, one
finds:
\begin{eqnarray} \label{eq:hosol}
\lefteqn{\int_0^t G^{-}_1(k,t-s) G^{+}_1(k,t-s) = }\nonumber
\\
& & \qquad = \sum_{\ell,\ell' = 2}^3
\frac{z_{\ell}z_{\ell'}}{(z_{\ell} +
z_{\ell'})(z_{\ell}-i\omega_k)(z_{\ell'}+i\omega_k)}\left[ \frac{z -
z_{\ell}}{H(z)}\right]_{z = z_{\ell}} \left[ \frac{z -
z_{\ell'}}{H(z)}\right]_{z = z_{\ell'}} \left[ e^{(z_{\ell} +
z_{\ell'})t} - 1 \right] \nonumber \\
& & \qquad  - i \sum_{\ell = 2}^3 \frac{\omega_k
z_{\ell}}{H(-i\omega_k)(z_{\ell} - i \omega_k)^2} \left[ \frac{z -
z_{\ell}}{H(z)}\right]_{z = z_{\ell}} \left[ e^{(z_{\ell} - i
\omega_k) t} - 1 \right] \nonumber \\
& & \qquad  + i \sum_{\ell = 2}^3 \frac{\omega_k
z_{\ell}}{H(+i\omega_k)(z_{\ell} + i \omega_k)^2} \left[ \frac{z -
z_{\ell}}{H(z)}\right]_{z = z_{\ell}} \left[ e^{(z_{\ell} + i
\omega_k) t} - 1 \right] \nonumber \\
& & \qquad + \frac{\omega_k^2}{H(-i\omega_k)H(+i\omega_k)}\, t.
\end{eqnarray}
The formula is rather cumbersome. However the terms in the first three lines
contain exponentially decaying terms, which vanish very rapidly with time. For
example---with reference to Eq.~\eqref{eq:sol}---the decay time is about $2.93
\times 10^{-47} \, {\rm s}$ for a 11 Kev photon. Accordingly, in the {\it large
time limit} we have for the differential photon emission rate $d \Gamma_k / dk$
(where, as in the free particle case, we have differentiated Eq.~\eqref{eq:ser}
over time, summed over the polarization states, and integrated over all
directions in photon's momentum space) the following simple {\it large-time}
expression:
\begin{equation} \label{eq;fddcxz}
\frac{d \Gamma_k}{dk} \; = \; \frac{\lambda \hbar c e^2}{2 \pi^2
\epsilon_0}\, \frac{k^3}{m^2(\omega_0^2 - c^2k^2)^2 + \beta^2
c^6k^6}.
\end{equation}
Two comments are at order. The first important thing one notices is that
Eq.~\eqref{eq;fddcxz} does {\it not} reduce to~\eqref{eq:wnfr} in the free
particle limit. The reason for this incongruence can be traced back to
Eq.~\eqref{eq:hosol}, according to which the free particle limit ($\omega_0
\rightarrow 0$) and the large time limit ($t \rightarrow + \infty$) do not
commute, as one can prove by direct calculation.  From the physical point of
view, the reason for the discrepancy is that, in the large time limit, the
particle has the chance to move far enough to feel the edges of the harmonic
potential, no matter how weak the potential is. This mean that the particle is
never really free, even in the limit $\omega_0 \rightarrow 0$. As a further
proof of this statement, one can note that by taking the free-particle limit at
{\it finite} times, one does indeed recover Eq.~\eqref{eq:wnfr}.

As a second observation, one can see that in the lowest order in the relevant
parameters ($\beta = 0$), the emission rate given by Eq.~\eqref{eq;fddcxz} is
of order $1/k$ for $ck \gg \omega_0$. This is reminiscent of the free particle
case. However, the exact expression is of order $1/k^3$, and the total emission
is finite, contrary to what implied by the free-particle expression. The
physical reason is that the binding potential works against the emission of
high-energy photons, as the term $e^{i {\bf k} \cdot {\bf x}}$ in
Eq.~\eqref{eq:A}, which is neglected by the dipole approximation, does.

The third relevant observation is that Eq.~\eqref{eq;fddcxz} shows a resonant
behavior in correspondence to the natural frequency $\omega_0$ of the
oscillator. Indeed the peak of the resonance is very high, due to the fact that
$\beta^2 c^6k^6$ is a very small quantity (confront the small value of $\beta$
given in Eq.~\eqref{eq:Hfun}) for $k = \omega_0/c$, where $\omega_0$ is a
standard frequency such as that associated to the hydrogen atom. Indeed such a
great resonance is incompatible with experimental data and, as such, it would
disprove this model, for any significant value of the collapse parameter
$\lambda$. However, the large value of the peak is an artificial feature of the
model. It emerges as a combination both of the fact that the the energy levels
of the harmonic oscillator are equally spaced, and from the dipole
approximation, according to which transitions are allowed only between to
consecutive levels. In order words, what happens here is that the noise excites
the particle to an higher energy level state; in the de-excitation process only
photons of energy $\hbar\omega_0$ can be emitted. In a more realistic model,
also photons with any energy $n\hbar\omega_0$ should be emitted, and the
spectrum would have a more articulated resonance structure, where the peaks are
less pronounced. An accurate spectrum would then display several resonances.

To conclude, our analysis shows that, in presence of a discrete spectrum (e.g.
the hydrogen atom), the differential photon emission rate due to the collapse
process should show a typical resonant behavior, which has not been depicted by
previous analysis. Although it is reasonable to expect that these resonances
are highly suppressed, it is worthwhile analyzing such a behavior for the CSL
model, by generalizing the previous results of~\cite{fu3,fu4} to the
low-frequency part of the spectrum.

\section{Conclusions}

We have analyzed the electromagnetic properties of both a free particle and of
a particle bounded by an harmonic potential, within the framework of collapse
models. By choosing a particularly simple, yet physically meaningful, model of
spontaneous wave function collapse, and under only the dipole approximation, we
have been able to solve the equations of motion exactly.

In the free particle case, we have found a counterintuitive result: the
particle's kinetic energy steadily increases in time, and at the same time it
spontaneously emits radiation at a constant rate. Although this is in principle
possible, as long as no conflict with experimental data emerges, such a
behavior suggests that collapse models should be modified in order the temper
(or eliminate entirely) the evident violation of the energy conservation
principle.

We have also found some discrepancies between our formula and those previously
derived, through a perturbative analysis. The origin of these differences is
not clear yet, and will be further studied in the future.

In the case of a particle confined by an harmonic potential, the spectrum is
modified and a peak emerges, in correspondence to the natural frequency of the
oscillator. This feature suggests that also in more realistic situations (e.g.
atomic systems) the spectrum should have a resonant structure, which is
worthwhile analyzing.

These results show that further analysis is required in order to better
understand the electromagnetic properties of charged particles in the CSL
model. This is important both for clarifying the theoretical picture offered by
collapse models, and also in the light of future experimental tests.

\section*{Acknowledgements}
We wish to thank S.L. Adler and G.C. Ghirardi for many useful conversations.
The work was supported by DFG, Germany. AB acknowledges the hospitality of the
Institute for Advanced Study of Princeton, were part of this work has been
done.

\begin{appendix}
\section{Derivation of Eq.~(31)}
The zeros of $H(z)$ defined by Eq.~\eqref{eq:Hfun} correspond to the solution
of the cubic equation:
\begin{equation}
z^3 + a_2 z^2 + a_1 z + a_0 = 0, \qquad\qquad a_0 =
-\frac{\kappa}{\beta}, \;\; a_1 = 0, \;\; a_2 = -\frac{m}{\beta}.
\end{equation}
By defining
\begin{equation}
q = \frac{1}{3}\, a_1 - \frac{1}{9}\, a_2^2\, \qquad r = \frac{1}{6}
\left( a_1 a_2 - 3 a_0 \right) - \frac{1}{27}\, a_2^3, \qquad
s_{1,2} = \sqrt[3]{r \pm \sqrt{q^3 + r^2}},
\end{equation}
the three roots can be written as:
\begin{equation} \label{eqa:roots}
z_1 = \left( s_1 + s_2 \right) - \frac{1}{3} a_2, \qquad z_{2,3} = -
\frac{1}{2}\, \left( s_1 + s_2 \right) - \frac{1}{3}\, a_2 \pm i
\frac{\sqrt{3}}{2} \left( s_1 - s_2 \right).
\end{equation}
This is the standard Cardan's method for finding the roots. In our case:
\begin{equation}
q^3 + r^2 \; = \; \frac{1}{4}\, a_0^2 \left( 1 + \frac{4}{27}\,
\frac{a_2^3}{a_0} \right) \; = \; \frac{1}{4} \,
\frac{\kappa^2}{\beta^2} \left( 1 + \frac{4}{27}\,
\frac{m^2}{\omega_0^2 \beta^2} \right) \; \simeq \; \frac{1}{27}\,
\frac{\omega_0^2 m^4}{\beta^4}
\end{equation}
if $\omega_0 \ll 2 m/ \sqrt{27} \beta$, as we have originally
assumed. Then $\sqrt{q^3 + r^2} \simeq \omega_0 m^2 / \sqrt{27}
\beta^2$. Working under the same approximation we have:
\begin{equation}
r \pm \sqrt{q^3 + r^2} \; \simeq \; \frac{1}{27} \frac{m^3}{\beta^3}
\left( 1 \pm \sqrt{27}\, \frac{\omega_0 \beta}{m} + \frac{27}{2}
\frac{\omega_0^2 \beta^2}{m^2} \right)
\end{equation}
and
\begin{equation}
s_{1,2} \; = \; \sqrt[3]{r \pm \sqrt{q^3 + r^2}} \; \simeq \;
\frac{1}{3} \frac{m}{\beta} \left(1 \pm \sqrt{3}\, \frac{\omega_0
\beta}{m} + \frac{3}{2} \frac{\omega_0^2 \beta^2}{m^2} \right).
\end{equation}
From the above expression and from Eq.~\eqref{eqa:roots}, the
approximate values of the roots given in~\eqref{eq:sol} can be
immediately derived.

\end{appendix}

\end{document}